\documentstyle[twocolumn,ucsd,epsf]{article}

\def\roughly#1{\raise.3ex
    \hbox{$#1$\kern-.75em\lower1ex\hbox{$\sim$}}}
\def\be{\begin{equation}}
\def\ee{\end{equation}}
\def\bea{\begin{eqnarray}}
\def\eea{\end{eqnarray}}

\newcommand{\square}{\kern1pt\vbox{\hrule height 0.6pt\hbox{\vrule width
0.6pt\hskip 3pt
   \vbox{\vskip 6pt}\hskip 3pt\vrule width 0.6pt}\hrule height 0.6pt}\kern1pt}

\newcommand{\<}{\!\!\!\!\!}

\begin{document}

\renewcommand{\thepage}{\roman{page}}
\renewcommand{\thebean}{\roman{bean}}
\setcounter{page}{-1}
\setcounter{bean}{0}

\begin{center}
March 1993\hfill   UCSD-PTH 93-07

\vskip .6in

\renewcommand{\thefootnote}{\fnsymbol{footnote}}
\setcounter{footnote}{0}

{\large\bf
The Higgs Model with a Complex Ghost Pair}
\footnote{ \noindent This work was supported by the
U.S. Department of Energy under Grant\newline
\indent\indent DE-FG03-91ER40546.}

\renewcommand{\thefootnote}{\alph{footnote}}
\setcounter{footnote}{0}

\vskip .4in

Karl Jansen,\footnote{E-mail: jansen@higgs.ucsd.edu}
Julius Kuti,\footnote{E-mail: kuti@sdphjk.ucsd.edu}
Chuan Liu\footnote{E-mail: chuan@higgs.ucsd.edu}

\vskip 0.2in

{\em Department of Physics 0319\\
     University of California at San Diego \\
        9500 Gilman Drive\\
        La Jolla, CA 92093-0319 USA}

\vskip 2cm

{\em Submitted to Physics Letters B}

\vskip 3cm

{\em Part One of Extended UCSD-PTH 92-40}

\end{center}

\newpage

\mbox{}

\vskip 0.7in

\begin{center}
{\bf Abstract}
\end{center}

\vskip 0.2in

\begin{small}
\begin{quotation}

A higher derivative term is introduced in the
kinetic energy of the Higgs Lagrangian in the minimal Standard Model.
A logically consistent and {\it finite} field theory is obtained
when some  excitations
of the Higgs field are quantized with indefinite metric in the
Hilbert space.
The Landau ghost phenomenon of the conventional triviality problem is
replaced by the state vectors of a complex ghost pair at a finite mass scale
with observable physical
consequences.
It is shown that
the ghost states exhibit unusual resonance properties and
correspond
to a complex conjugate pair of Pauli-Villars regulator masses in the
euclidean
path integral formulation of the theory.
An argument is given that microscopic acausality effects associated with the
ghost pair
remain undetectable in scattering processes with realistic wave
packects, and the
S-matrix should exhibit unitarity in the observable sector of the Hilbert
space.
\end{quotation}
\end{small}

\newpage



\renewcommand{\thepage}{\arabic{page}}
\renewcommand{\thebean}{\arabic{bean}}
\setcounter{page}{0}
\setcounter{bean}{0}

\renewcommand{\thefootnote}{\arabic{footnote}}
\setcounter{footnote}{0}


\section{Introduction}

We introduce a higher derivative term in the
kinetic energy of the Higgs Lagrangian to
keep quantum fluctuations finite while
all the symmetries of the model are preserved. An
unconventional viewpoint is taken to
investigate the higher derivative Lagrangian as a {\it finite} field theory
with calculable scattering amplitudes at arbitrary energies.
The Landau ghost phenomenon of the triviality problem \cite{LANDAU}
is  replaced by a genuine complex ghost pair with observable physical
consequences.
We will argue that this unusual scenario
is not excluded logically, or experimentally,
in the Higgs sector of the Standard Model.
The new Lagrangian offers a calculable strongly interacting Higgs sector with
a heavy Higgs particle in the TeV region without internal inconsistencies,
or physically implausible cut-off effects.
Although many aspects of our work
are generally applicable to higher derivative field theories,
the Higgs model is presented as our first illustration of the approach.
In the future we plan to investigate
higher derivative
gauge field models with fermions, and higher derivative quantum gravity
\cite{GRAVITY}.

Our
approach is in the spirit of an early proposal by Podolsky \cite{POD}
who rendered the electron self-energy problem finite in
quantum electrodynamics by adding
a higher derivative regulator term to the Maxwell Lagrangian. The new term
changes the physics of electromagnetism at short distances. The connection
between the higher derivative
Maxwell theory and Pauli-Villars regulators \cite{VILLARS}
had been pointed out \cite{PAIS}.
The
idea was later revisited  by Lee and
Wick \cite{LEE} who did not start from the higher
derivative Maxwell Lagrangian, instead they introduced an auxiliary
heavy photon
field which was quantized with indefinite metric. They emphasized the logical
and experimental possibility of a heavy photon represented by
a complex ghost pole in electron-positron scattering amplitudes. The arguments
of Lee and Wick were based on modified rules of contour integration in the
evaluation of  Feynman diagrams and
a non-perturbative path
integral formulation remained elusive \cite{GROSS}.

We present new results here on the quantization of higher derivative
Lagrangians with indefinite metric in the Hilbert space. It will be shown
that in a scalar field theory with $\Phi^4$ interaction the higher
derivative Lagrangian and its quantization with indefinite metric leads
to a consistent
non-perturbative approach. The euclidean path integral will be derived from
the Hamiltonian formalism and the origin of the conventional Pauli-Villars
regulator will be clarified. The phenomenological implications of a complex
ghost pair will be briefly described.

\section{Higher Derivative Quantum Mechanics}

A simple quantum mechanical model will illustrate
the strategy of our calculations.
Consider the classical
Lagrangian of a non-relativistic particle moving in one dimension:
\begin{equation}
L= \frac{1}{2}\Bigl(1\!+\!\frac{2\omega^2}{M^2}
{\rm cos}2\Theta\Bigr)\dot x^2\!
    -\!\frac{1}{M^2}\Bigl({\rm cos}2\Theta\!+\!\frac{\omega^2}{2M^2}\Bigr)\ddot
x^2  
    + \frac{1}{2M^4}{\stackrel{\textstyle ...}{x}}^2 - \frac{1}{2}\omega^2x^2
\;.
\label{eq:QMLag}
\end{equation}
Eq.~\ref{eq:QMLag} describes a simple harmonic oscillator with second
and third derivative terms added to the original Lagrangian
whose potential energy term is $V(x)=1/2~\omega^2x^2$;
$M$ and $\omega$ are measured
in units of the Newton mass $m$ of the original oscillator particle, set
to $m=1$ for convenience.
For simple interpretation of the results, the
coefficients of the derivative terms are given in terms of $M$ and an angle
parameter $\Theta$;
the only restrictions imposed are $\omega^2/M^2
< 1$ and $0 \! < \! \Theta \! < \! \pi/2$.

The variational principle
leads to the higher order Euler-Lagrange equation of motion,
\begin{equation}
\frac{1}{M^4}\frac{d^6 x}{dt^6} +
\frac{2}{M^2}\Bigl({\rm
cos}2\Theta\!+\!\frac{\omega^2}{2M^2}\Bigr)\frac{d^4x}{dt^4} ~-
  \Bigl(1\!+\!\frac{2\omega^2}{M^2}
{\rm cos}2\Theta\Bigr)\frac{d^2x}{dt^2} + \omega^2 x = 0 ~,
\label{eq:EulLag}
\end{equation}
which differs from Newton's equation by higher derivative terms with small
coefficients in the limit $\omega^2/M^2 \ll 1$, and $1/M^2 \ll 1$
(limit of large regulator masses in field theory).
Even very small higher derivative
terms will have a qualitative impact on the theory, as shown in the following
analysis.

The canonical formalism for higher derivative theories was developed first
by Ostrogradski \cite{OSTRO,PAIS}. Three independent generalized coordinates
$x$, $\dot x$, and $\ddot x$  are introduced and the
generalized canonical momenta are defined by
\begin{eqnarray}
\< \pi_x \< &=& \< \frac{\partial L}{\partial\dot x} -
\frac{d}{dt}  \frac{\partial L}{\partial \ddot x} +
\frac{d^2}{dt^2}\frac{\partial L} {\partial {\stackrel{\textstyle ...}{x}}}
= \nonumber \\
& & \!\! \Bigl(1\!+\!\frac{2\omega^2}{M^2}{\rm cos}2\Theta\Bigr)\dot x
+
\frac{2}{M^2}\Bigl({\rm
cos}2\Theta\!+\!\frac{\omega^2}{2M^2}\Bigr){\stackrel{\textstyle ...}{x}}
+ \frac{1}{M^4}\frac{d^5x}{dt^5}\; , \nonumber \\
\nonumber \\
\< \pi_{\dot x}\< &=&\< \frac{\partial L}{\partial\ddot x}\! -\!
\frac{d}{dt}  \frac{\partial L}{\partial {\stackrel{\textstyle ...}{x}}} =
-  \Bigl(\frac{2{\rm cos}2\Theta}{M^2} + \frac{\omega^2}{M^4}\Bigr)\ddot x
- \frac{1}{M^4}\frac{d^4x}{dt^4}, \nonumber \\ \nonumber \\
\pi_{\ddot x}\< &=& \frac{\partial L}{\partial {\stackrel{\textstyle
...}{x}}} = \frac{1}{M^4}{\stackrel{\textstyle ...}{x}} \; .
\label{eq:can_mom}
\end{eqnarray}
With Ostrogradski's method we find the Hamiltonian
\begin{eqnarray}
\lefteqn{
\<\<\< H  = \pi_x\cdot\dot x + \pi_{\dot x}\cdot\ddot x +
\frac{1}{2}M^4\pi_{\ddot x}^2 + \frac{1}{2}\omega^2 x^2
}
\nonumber \\
& &\mbox{}
-\frac{1}{2}\Bigl(1+2\frac{\omega^2}{M^2}{\rm cos}2\Theta\Bigr)\dot x^2 +
\frac{1}{2}\Bigl(\frac{\omega^2}{M^4}+\frac{2{\rm cos}2\Theta}{M^2}\Bigr)\ddot
x^2  . \label{eq:hamiltonian}
\end{eqnarray}
In quantum mechanics the operators of the canonical variables satisfy
standard Heisenberg commutation relations,
\begin{equation}
[ \hat{x}, \hat{\pi}_x ] = i\hbar ~, ~~~~~
[ \hat{\dot x} , \hat{\pi}_{\dot x} ] = i\hbar ~, ~~~~~
[ \hat{\ddot x} , \hat{\pi}_{\ddot x} ] = i\hbar \; .
\label{eq:COMMUTATOR}
\end{equation}
All the other commutators of the six operators vanish. We define
the Hilbert
space with indefinite metric \cite{PAULI}
by the
scalar product
\begin{equation}
\langle x', \dot x', \ddot x' | x, \dot x, \ddot x \rangle =
\delta (x'-x)~\delta (\dot x' + \dot x)~\delta (\ddot x'-\ddot x) ~,
\label{eq:metric}
\end{equation}
as indicated by the {\it plus sign} in the second $\delta$-function on the
right side of Eq.~\ref{eq:metric}.
The wave function $\psi(x,\dot x, \ddot x) =$
$\langle x, -\dot x, \ddot x |\psi\rangle$
depends on the three independent generalized coordinates.
The completeness relation in the indefinite metric Hilbert space is given by
\begin{equation}
\int\!\!\! dx\!\!\! \int \!\!\! d\dot x\!\!\! \int\!\!\! d\ddot x
\; |x, -\dot x, \ddot x\rangle\langle x, \dot x, \ddot x | = {\rm I} \; .
\end{equation}
The representation of the self-adjoint operators $\hat{x}, \hat{\pi}_x$,
$\hat{\ddot x},  \hat{\pi}_{\ddot x}$
follows the rules of ordinary quantum mechanics,
$\hat{x}~\psi = x\cdot\psi$, $\hat{\pi}_x~\psi =
-i\hbar~\frac{\partial\psi}{\partial x}$, and similarly,
$\hat{\ddot x}~\psi = \ddot x\cdot\psi$, $\hat{\pi}_{\ddot x}~\psi =
-i\hbar~\frac{\partial\psi}{\partial \ddot x}$.
However, $\hat{\dot x}$ and $\hat{\pi}_{\dot x}$
(also self-adjoint with respect to the indefinite metric)
are represented by
$\hat{\dot x}~\psi = i\dot x\cdot\psi$, and $\hat{\pi}_{\dot x}~\psi =
-\hbar \frac{\partial\psi}{\partial \dot x}$.

The euclidean partition function in the presence of external sources is defined
by
\begin{eqnarray}
Z_T(J_0,J_1,J_2) &=& {\rm Tr} e^{-\frac{1}{\hbar} T\, H}  \nonumber \\
&=& \int\!\!\! dx\!\!\! \int \!\!\! d\dot x\!\!\!
\int\!\!\! d\ddot x \; \langle x, -\dot x, \ddot x|
e^{-\frac{1}{\hbar} T\, H}|x, \dot x, \ddot x \rangle \; ,
\label{eq:partf}
\end{eqnarray}
where a time-dependent source term $J_0\hat{x} + J_1\hat{\dot x} + J_2\hat{
\ddot x}$ is included in the Hamiltonian.
The partition function of Eq.~\ref{eq:partf} can be written as a Hamiltonian
path integral in six canonical variables. After integrating over
$\dot x, \ddot x, \pi_x, \pi_{\dot x}, \pi_{\ddot x}$, we find
\begin{eqnarray}
\<\<\<\<\<& &\<\< Z_T(J_0, J_1, J_2) = \int d[x(\tau)]~{\rm exp}
\Biggl\{ -\frac{1}
{\hbar} \int^T_0 d\tau \biggl(
L_E + J_0(\tau) x(\tau)  \nonumber \\
& & \mbox{} ~~~~~~~~ + J_1(\tau)\frac{dx}{d\tau} +
J_2(\tau)\frac{d^2x}{d\tau^2}\; \biggr) \Biggr\} ,
\label{eq:ZT}
\end{eqnarray}
where $L_E$ is the {\it euclidean} Lagrangian of the higher derivative model,
\begin{eqnarray}
\lefteqn{
\<\<\<\< L_E = \Bigl( \frac{1}{2} + \frac{\omega^2}{M^2}{\rm cos}
2\Theta\Bigr)\Bigl( \frac{dx}{d\tau}\Bigr)^2 + \frac{1}{2}\omega^2x^2(\tau)
} \nonumber \\
& &\mbox{} +
\frac{1}{M^2}\Bigl( {\rm cos}2\Theta +
\frac{\omega^2}{2M^2}\Bigr)\Bigl(\frac{d^2x}{d\tau^2}\Bigr)^2  +
\frac{1}{2}M^{-4}\Bigl(\frac{d^3x}{d\tau^3}\Bigr)^2 ~.
\label{eq:LE}
\end{eqnarray}
In the T $\rightarrow \infty$ limit, after the evaluation of the path integral
in Eq.~\ref{eq:ZT}, we obtain
\begin{eqnarray}
\lefteqn{
\<\<\<\< Z_{\infty}(\tilde{J}_0(E), \tilde{J}_1(E), \tilde{J}_2(E)) =
Z_{\infty}(0,0,0)
} \nonumber \\
 &\times&
{\rm exp} \Biggl\{ -\frac{1}{2} \int dE\tilde{J}(E)\widetilde{D}(E)
\tilde{J}(-E) \Biggr\} ~ ,
\label{eq:Zinf}
\end{eqnarray}
where $\tilde{J}(E)=\tilde{J}_0(E) - i E\tilde{J}_1(E) - E^2\tilde{J}_2(E)$.
The Fourier transform of the quantum mechanical propagator $D(\tau)$ is
given by
\begin{equation}
\widetilde{D}(E) =
\frac{M^4}{(E^2+\omega^2)(E^2+M^2e^{2i\Theta})(E^2+M^2e^{-2i\Theta})}~~.
\label{eq:D(E)}
\end{equation}

The physical significance of the multiple pole structure in the
propagator becomes clear
from the spectrum of the Hamiltonian which can be
transformed to a diagonal form,
\begin{equation}
H =
\hbar\Bigl[\omega a^{(+)}a^{(-)} + Me^{i\Theta}b^{(+)}b^{(-)}
+  Me^{-i\Theta}c^{(+)}c^{(-)}
 + \! \frac{1}{2} (\omega +  M e^{i\Theta} +
 M e^{-i\Theta})\Bigr] \; .
\end{equation}
The creation operators $a^{(+)}, b^{(+)}, c^{(+)}$ and the annihilation
operators $a^{(-)}, b^{(-)}, c^{(-)}$ are linear combinations
of the six canonical variables $x, \dot x, \ddot x, \pi_x , \pi_{\dot x},
\pi_{\ddot x}$ with commutation relations
$[ a^{(-)}, a^{(+)} ] =1$, $[ b^{(-)}, b^{(+)} ] =1$,
$[ c^{(-)}, c^{(+)} ] =1$.
All the other commutators vanish. The adjoint
of $a^{(+)}$ is $\overline{a^{(+)}} = a^{(-)}$, and similarly,
$\overline{b^{(+)}} =
c^{(-)}$, $\overline{c^{(+)}} = b^{(-)}$.

The eigenstates of the Hamiltonian are given by
\begin{equation}
|l,m,n \rangle =
\frac{1}{\sqrt{n!}}(c^{(+)})^n\frac{1}{\sqrt{m!}}(b^{(+)})^m
\frac{1}{\sqrt{l!}}(a^{(+)})^l\; |0,0,0\rangle \; ,
\end{equation}
where the ground state $|0,0,0\rangle$ is annihilated by $a^{(-)}$,
$b^{(-)}$, $c^{(-)}$. The complex energy eigenvalues
\begin{eqnarray}
E_{l,m,n} \!=\! ( l + \frac{1}{2} )\hbar\omega + ( m + \frac{1}{2} )
\hbar Me^{i\Theta} \!+\! ( n + \frac{1}{2} ) \hbar Me^{-i\Theta}
\end{eqnarray}
satisfy the symmetry relation $E_{l,m,n} = E^{*}_{l,n,m}$ . In addition
to the ordinary oscillator
excitations we have two conjugate ghost oscillators with complex energies,
in agreement with the two new degrees of freedom in the Lagrangian.
The energy eigenstates have indefinite metric normalization,
\begin{equation}
\langle l',m',n' |l,m,n \rangle = \delta_{l'l}\delta_{m'n}\delta_{n'm} \; .
\label{eq:norm}
\end{equation}
For $m=n$, the ghost oscillator and its complex conjugate partner are excited
in pairs and the corresponding state vectors have positive norm with real
energy. However, unbalanced ghost
excitations
with $m\neq n$ correspond to zero norm state vectors with complex energies.
Some linear combinations of ghost excitations with $m\neq n$ will have negative
norm.
%
%

The quantization procedure we developed has several important consequences.
The euclidean propagator $\widetilde{D}(E)$ of Eq.~\ref{eq:D(E)} can be written
as a sum of three simple pole terms describing the original oscillator and the
complex conjugate pair of ghost oscillators. The conjugate ghost poles play
the same role as the auxiliary Pauli-Villars regulators in field theory. The
method which is applicable for any potential energy $V(x)$ in the Lagrangian of
Eq.~\ref{eq:QMLag} leads to the euclidean path integral formulation of
Eq.~\ref{eq:ZT} providing a non-perturbative framework for numerical
investigations. The euclidean path integral, however, cannot be continued
to real time since the Wick rotation of the integration contour is blocked
by the complex ghost poles on the first sheet of the complex energy plane.
The ghost eigenstates with complex energies represent run-away time
evolution in the Hilbert space rendering the real-time path
integral ill-defined.
At the same time, the hamiltonian approach remains well-defined in
the indefinite metric Hilbert space and provides the
theoretical foundation of the model in real time.
Important informations can also be extracted
directly from the euclidean path integral of Eq.~\ref{eq:ZT}.


%
\section{Higher Derivative Higgs Model}

In the minimal Standard Model with $SU(2)_L \times U(1)$ gauge
symmetry the Higgs sector is described by a complex
scalar doublet $\Phi$ with quartic self-interaction.
The Higgs potential has the well-known form
$
V(\Phi^{\dagger}\Phi) = - {1\over2} m_0^2 \Phi^{\dagger}\Phi +
\lambda_0 (\Phi^{\dagger}\Phi)^2 ~
$
where $m_0^2$ is a mass parameter ($m_H=\sqrt 2 m_0$ in tree approximation)
and $\lambda_0$ designates the quartic coupling
constant.
The four real components of the complex scalar doublet $\Phi$ transform
as a real vector $\phi_\alpha, ~\alpha=1,2,3,4$,
under O(4) symmetry transformations while  the Higgs potential remains
invariant.
In the limit where the gauge couplings and Yukawa couplings are neglected
the quantization procedure for the higher derivative Higgs Lagrangian,
\begin{equation}
L = \frac{1}{2} \Bigl( \partial_\mu \vec\phi\partial^\mu\vec\phi
+ m^2_0\vec\phi\cdot\vec\phi \Bigr)
 + \frac{1}{2M^4}~\square\partial_\mu\vec\phi~\square\partial^\mu\vec\phi
- \lambda_0\Bigl( \vec\phi\cdot\vec\phi\Bigr) ^2 ~,
\label{eq:O(4)_Lagr}
\end{equation}
in Minkowski space-time is very similar to the earlier quantum mechanical
example.
There are three independent sets of field variables (generalized coordinates)
$\phi_\alpha (\vec x, t)$, $\dot{\phi}_\alpha (\vec x, t)$, and
$\ddot{\phi}_\alpha (\vec x, t)$, $\alpha=1,2,3,4$. The canonical
field momentum variables $\Pi_{\phi_\alpha}$, $\Pi_{\dot{\phi}_\alpha}$,
$\Pi_{\ddot{\phi}_\alpha}$ are defined in close analogy with
Eq.~\ref{eq:can_mom}.
The Heisenberg commutators of the canonical
variables have the same form as in ordinary quantum field theory with
positive definite metric, in accord with Eq.~\ref{eq:COMMUTATOR}.
The state vectors $|\psi\rangle$ are described by wave functionals
in the field-diagonal representation of the indefinite metric Hilbert space,
\begin{equation}
\psi\Bigl\{ \phi_\alpha (\vec x), \dot{\phi}_\alpha (\vec x),
\ddot{\phi}_\alpha (\vec x) \Bigr\} =
\langle \phi_\alpha (\vec x), - \dot{\phi}_\alpha (\vec x),
\ddot{\phi}_\alpha (\vec x) | \psi \rangle ~.
\end{equation}
The operators $\hat{\phi}_\alpha$, $\hat{\Pi}_{\phi_\alpha}$
are represented in the Hilbert space conventionally,
\begin{equation}
\hat{\phi}_\alpha (\vec x)~\psi\Bigl\{ \phi_\alpha (\vec x),
\dot{\phi}_\alpha (\vec x),
\ddot{\phi}_\alpha (\vec x) \Bigr\} =
{}~~~~~\phi_\alpha (\vec x)\cdot\psi\Bigl\{ \phi_\alpha (\vec x),
\dot{\phi}_\alpha (\vec x),
\ddot{\phi}_\alpha (\vec x) \Bigr\},
\end{equation}
\begin{equation}
\hat{\Pi}_{\phi_\alpha}~\psi\Bigl\{ \phi_\alpha (\vec x),
\dot{\phi}_\alpha (\vec x),
\ddot{\phi}_\alpha (\vec x) \Bigr\} =
- i\hbar \frac{\delta}{\delta \phi_\alpha (\vec x)} ~
\psi\Bigl\{ \phi_\alpha (\vec x), \dot{\phi}_\alpha (\vec x),
\ddot{\phi}_\alpha (\vec x) \Bigr\},
\end{equation}
with similar representation for the
$\hat{\ddot{\phi}}_\alpha$, $\hat{\Pi}_{\ddot{\phi}_\alpha}$ pair.
However, the representation of the
$\hat{\dot{\phi}}_\alpha$, $\hat{\Pi}_{\dot{\phi}_\alpha}$ pair reveals the
indefinite metric:
\begin{equation}
\hat{\dot{\phi}}_\alpha (\vec x)~\psi\Bigl\{ \phi_\alpha
(\vec x), \dot{\phi}_\alpha (\vec x),
\ddot{\phi}_\alpha (\vec x) \Bigr\} =
i~\dot{\phi}_\alpha (\vec x)\cdot\psi\Bigl\{ \phi_\alpha (\vec x),
\dot{\phi}_\alpha (\vec x),
\ddot{\phi}_\alpha (\vec x) \Bigr\} ~,
\end{equation}
\begin{equation}
\hat{\Pi}_{\dot{\phi}_\alpha}~\psi\Bigl\{ \phi_\alpha (\vec
x), \dot{\phi}_\alpha (\vec x),
\ddot{\phi}_\alpha (\vec x) \Bigr\} =
- \hbar \frac{\delta}{\delta \dot{\phi}_\alpha (\vec x)} ~
\psi\Bigl\{ \phi_\alpha (\vec x), \dot{\phi}_\alpha (\vec x),
\ddot{\phi}_\alpha (\vec x) \Bigr\} ~.
\end{equation}

Following the canonical procedure of the quantum mechanical model, the Hamilton
operator can be expressed in terms of the field variables
$\phi_\alpha (\vec x)$, $\dot{\phi}_\alpha (\vec x)$,
$\ddot{\phi}_\alpha (\vec x)$ and the canonical
field momentum variables $\Pi_{\phi_\alpha}$, $\Pi_{\dot{\phi}_\alpha}$,
$\Pi_{\ddot{\phi}_\alpha}$.
The euclidean partition function
$Z_T = {\rm Tr} e^{-\frac{1}{\hbar} T\, H}$ is represented by
a Hamiltonian path integral in the six canonical variables.
It is straightforward to derive the conventional euclidean path integral by
integrating over the variables
$\dot{\phi}_\alpha$,
$\ddot{\phi}_\alpha$,
$\Pi_{\phi_\alpha}$, $\Pi_{\dot{\phi}_\alpha}$,
$\Pi_{\ddot{\phi}_\alpha}$. In the $T \to \infty$ limit we find
\begin{equation}
         Z = \int [d\vec\phi]~{\rm exp} ~ \Bigl\{ -
\frac{1}{\hbar}S_E[\vec\phi] \Bigr\}\; ,
\label{eq:ZE1}
\end{equation}
where the euclidean action $S_E$ is given by
\begin{equation}
S_E  = \int d^4x \Bigl[ - {1\over2} ~
\vec\phi~(\square + {1\over M^4}\square^3)~\vec\phi
- {1\over2} m_0^2~\vec\phi\cdot\vec\phi + {\lambda_0}
(\vec\phi\cdot\vec\phi)^2 ~ \Bigr] ~ ,
\label{eq:SEO4}
\end{equation}
and $\square$ is the
euclidean Laplace operator in four dimensions.
The higher
derivative term $M^{-4}\square^3$  acts
in the inverse euclidean propagator as a
Pauli-Villars regulator with mass parameter $M$. It represents a minimal
modification of the continuum model,
if we want to render the field theory and its euclidean path integral finite.
Like in higher derivative quantum mechanics, the euclidean propagator is
the sum of three simple poles: a complex conjugate pair of ghost poles is
added to the simple pole representing the original O(4) particle.

The model defined by the partition function of Eq.~\ref{eq:ZE1} has
two phases, as expected.
In the symmetric phase
we find the original massive particle (with four components in the
intrinsic O(4) space), and we also find a complex ghost pair
with intrinsic
O(4) symmetry whose mass scale is set by the Pauli-Villars
mass parameter $M$.
The general
particle content of the model is exhibited in the symmetric phase by the
Hamiltonian in terms of creation and annihilation operators,
$H=H_0 + H_{int}$,
\begin{equation}
 H_0 = \sum\hbar \Bigl[ \omega(\vec p) a^{(+)}_\alpha (\vec p) a^{(-)}_\alpha
(\vec p) + \Omega(\vec p) b^{(+)}_\alpha (\vec p) b^{(-)}_\alpha (\vec p)
+ \Omega^{*}(\vec p) c^{(+)}_\alpha (\vec p) c^{(-)}_\alpha (\vec p) \Bigr]~,
\label{eq:HAM}
\end{equation}
where the summation is over internal O(4) indices and momentum modes.
The dispersion of the massive O(4) particle is given by $\omega(\vec p) =
\sqrt{\vec p^2 + m^2}$ whereas the complex energy dispersion of the ghost
state is  $\Omega(\vec p) = \sqrt{\vec p^2 + {\cal M}^2}$,
with ${\cal M} = M\cdot e^{i\Theta}$. In a more conventional parametrization
one separates the real and imaginary parts,  ${\cal M}=M_G + i\gamma/2$,
$M_G=M{\rm cos}\Theta,~\gamma =2M{\rm sin}\Theta$.
In our Lagrangian the complex
phase $\Theta$ is a function of $m$ and $M$. With the addition of a
$\square^2$ term to the Lagrangian the phase angle $\Theta$ can be promoted to
an independent parameter moving the ghost pair on the complex energy plane
without restrictions.
We will assume
${\rm tan}\Theta\approx 1$, so that $\gamma$ is of the same order of magnitude
as $M_G$ in our applications.

The operator $a_\alpha^{(+)}(\vec p)$ creates the Higgs
particle with momentum $\vec p$ and energy $\hbar\omega(\vec p)$. The operator
$b_\alpha^{(+)}(\vec p)$ creates a ghost state with momentum $\vec p$ and
complex energy $\hbar\Omega(\vec p)$; $c_\alpha^{(+)}(\vec p)$  creates the
complex conjugate ghost state. The commutation relations of the creation and
annihilation operators, their conjugation properties, and the metric properties
of the states they create are obtained in full analogy with the higher
derivative quantum mechanical oscillator.
The interaction part of the Hamiltonian is
proportional to the coupling constant and expressed in terms of creation and
annihilation operators.

In the broken phase we find a Higgs
particle with mass $m_H$, and three massless Goldstone excitations with
residual
O(3) symmetry. In addition,
the 4-component heavy ghost particle and its complex conjugate partner
also appear in the particle spectrum of the
broken phase.  One ghost state has the internal quantum
numbers of the Higgs particle. The other three ghost states carry the internal
quantum numbers of the Goldstone particles.

The hamiltonian formulation and the euclidean path integral
framework we derived
provide self-contained non-perturbative tools for further investigations.
In earlier work
Slavnov introduced
a formal path integral in Minkowski space-time, based on a canonical
quantization
procedure \cite{SLAVNOV}. The complex ghost states with run-away time
evolution render the
Minkowski path integral ill-defined and inadequate for
non-perturbative analysis.
Hawking, on the other hand, based his work \cite{HAWKING} on the euclidean
path integral approach as an
intuitive starting point without establishing the connection with the canonical
quantization procedure .

\section{The Implications of Ghost States}

The Hilbert space of the Higgs model with a complex ghost pair is divided into
two distinct parts. In the physical subspace the norm is positive and the
energy
eigenvalues of the stationary eigenvectors are real in {\it every} Lorentz
frame. Physical wave packets in observable initial and final states of
scattering processes are formed from the Hamiltonian eigenstates of the
physical
subspace. Ghost states with complex energy and non-positive norm are in
the unobservable nonphysical part of the Hilbert space. Although complex
conjugate ghost  pairs can have positive norm and real total energy in some
special Lorentz frame, their energy is complex in general Lorentz frames.
Accordingly, conjugate ghost pairs
are described by state vectors in the unobservable part of the Hilbert space.

The scattering amplitudes of physical states have
unusual singularity structures. Consider the elastic scattering of two
Goldstone
particles ($W_LW_L$ scattering via the equivalence theorem \cite{EQUIVALENCE}).
The scattering amplitude has a cut in the invariant
energy variable $s=(p_1 + p_2)^2$
along the real axis and two complex poles at $s = {\cal M}^2$ and $s =
{{\cal M}^*}^2$ on the physical sheet (the Higgs particle appears as an
ordinary resonance on the second sheet). The expected time advancement effects
of the complex ghost pole on the outgoing wave packet can be estimated. The
normalized
incident wave packet $\chi^{in}$ is given by $|\chi^{in}|^2 \sim
\frac{\Delta}{r^2} e^{-\Delta |r+t|}$ where $r$ designates the distance
between the colliding particles in the center of mass system and $\Delta$ is
the momentum width of the incident wave packet.
Assuming that $\Delta
>>\gamma$ the ghost pole contribution to the outgoing wave packet
$\chi^{out}$ has the intensity $|\chi^{out}|^2 \sim
\frac{\gamma^2}{\Delta r^2}e^{-\gamma|r-t|}$ for $r-t >> \Delta^{-1}$.
Since $t=0$ is the time of collision,
the intermediate ghost state acts in the scattering
process as a peculiar resonance with time advancement in the emergence of the
decay products. The scale of the time advancement is set by $\gamma$
and remains undetectable with realistic wave packets an experimentalist can
prepare in the initial state. In the experimental mass distribution of the
decay
products the ghost will appear as a resonance state of approximate mass
$M_G$.

In the more complicated intermediate state of Fig.~1a the
unphysical complex conjugate ghost pair generates a non-analytic singularity
in the physical region at
$s=({\cal M} + {\cal M}^*)^2$ which is a real physical value of the
invariant energy variable in the elastic scattering amplitude.
It can be shown that this diagram is compatible with unitarity because the
imaginary part of the corresponding scattering amplitude vanishes. Although the
ghost pair does not appear in the physical final state, the ghost loop
contributes to the real part of elastic scattering. The ghost loop also
contributes to the renormalization group $\beta$-function in a non-trivial way.
Cancellation effects between the ghost loop and ordinary particle loops will
determine the evolution of the effective  coupling constant of the scattering
process at asymptotic energies.
As a result, in the parameter regime of a strongly interacting Higgs sector,
the elastic $W_LW_L$ cross section can reach
the unitarity bound without Landau ghost problems.

In the inelastic channel, schematically depicted in Fig.~1b, the mass
distribution of the four Goldstone particles in the final state will show a
pair of resonance bumps, although the ghost pair does not propagate in the
intermediate state with macrosopic acausality effects.
We estimated that the time advancement effects of the
intermediate ghost pair in the outgoing scattered
wave of elastic scattering, or in the decay products of the inelastic channel
are too small to observe.


The non-perturbative computer investigation of the higher derivative Lagrangian
requires the introduction of an underlying hypercubic lattice structure:
the continuum operators $\square$ and $\square ^3$ are replaced in
Eq.~\ref{eq:SEO4} by the equivalent hypercubic lattice operators. The
lattice spacing $a$ defines a new short distance scale in the theory with the
associated lattice momentum cut-off at $\Lambda = \pi/a$.
At fixed $M/m_H$,
the limit $a \to 0$
corresponds to the finite higher derivative theory in the continuum.
The phenomenological implications of the strongly interacting Higgs sector
with a Higgs mass on the TeV mass scale and a ghost pair in the multi-TeV
energy range are
calculable in our model and we hope to return to quantitative details of this
interesting problem in future publications.

\subsection*{Acknowledgements}

We thank Aneesh Manohar and other members of the Particle Theory Group
at U.C. San Diego for useful conversations and comments. Critical remarks by
Howard Georgi and a referee are also
acknowledged. This work was supported by the DOE under Grant DE-FG03-91ER40546.


\newpage

\section*{Figure Captions}

\begin{description}

\vskip 1cm

\item[Fig.~1:]
(a) The simplest Feynman diagram that leads to non-analyticity.
Internal dotted lines are ghost propagators with complex poles, and the
solid line designates massless Goldstone particles.
(b) Inelastic production of Goldstone
particles through an intermediate ghost pair.

\end{description}

\vfill
\pagebreak

\begin{figure}[t]
\centerline{\epsfbox{07_fig1.ps}}
\end{figure}




\vskip 1cm

\begin{thebibliography}{99}
\bibitem{LANDAU}      L.~D.~Landau, A.~A.~Abrikosov, and I.~M.~Khalatnikov,
                      Doklady Akad. Nauk. USSR, 95 (1954) 1177;
                      L.~Maiani, G.~Parisi, R.~Petronzio, Nucl. Phys. B136
                      (1978) 115;
                      R.~Dashen and H.~Neuberger, Phys. Rev. Lett. 50
                      (1983) 1897.
\bibitem{GRAVITY}     K.~S.~Stelle, Phys. Rev. D16 (1977) 953;
                      E.~Tomboulis, Phys. Lett. 97B (1980)
                      77.
\bibitem{POD}         B.~Podolski, Phys. Rev. 62
                      (1942) 68; B.~Podolski and
                      P.~Schwed, Rev. Mod. Phys. 20 (1948) 40.
\bibitem{VILLARS}     W.~Pauli and F.~Villars, Rev. Mod. Phys. 21 (1949) 434.
\bibitem{PAIS}        A.~Pais and G.~E.~Uhlenbeck, Phys. Rev. 79 (1950) 145.
\bibitem{LEE}         T.~D.~Lee and G.~C.~Wick, Nucl. Phys. B9 (1968) 209;
                      Phys. Rev. D2 (1970) 1033.
\bibitem{GROSS}       D.~G.~Boulware and D.~J.~Gross, Nucl. Phys. B233 (1983)
                      1.
\bibitem{OSTRO}       M.~Ostrogradski, Mem. Ac. St. Petersbourg, 4(1850)
                      385.
\bibitem{PAULI}       W.~Pauli, Rev. Mod. Phys. 15 (175) 1943.
\bibitem{SLAVNOV}     A.~A.~Slavnov, Nucl. Phys. B31 (1971) 301.
\bibitem{HAWKING}     S.~W.~Hawking, Quantum Field Theory and Quantum
                      Statistics, edited by I.~A.~Batalin et al. (1987) 129.
\bibitem{EQUIVALENCE} B.~W.~Lee, C.~Quigg and H.~B.~Thacker, Phys.
                      Rev. D16 (1977) 1519.
\end{thebibliography}
\end{document}